\documentclass[twocolumn,showpacs,amsmath,amssymb,superscriptaddress,aps,prl]{revtex4-1}

\usepackage{ifpdf}
\usepackage{graphicx}
\usepackage{dcolumn}
\usepackage{bm}
\usepackage[dvipsnames]{xcolor}
\usepackage{subfigure}
\usepackage{xspace}
\usepackage{tabularx}
\usepackage{mathtools}

\newcommand{\tcb}[1]{\textcolor{black}{#1}}
\newcommand{\er}[1]{\textcolor{black}{#1}}
\newcommand{\tk}[1]{\textcolor{black}{#1}}
\newcommand{\mk}[1]{\textcolor{black}{#1}}
\newcolumntype{R}{>{\raggedleft\arraybackslash}X}%
\newcolumntype{L}{>{\raggedright\arraybackslash}X}%
\usepackage[version-1-compatibility,obeyall,per=slash,tophrase={-},parse-numbers=false]{siunitx}
\usepackage[version=3]{mhchem} 
\DeclareGraphicsExtensions{.pdf,.png,.jpg}

\ifpdf
  \graphicspath{{figures/pdf/}}
\else
  \graphicspath{{figures/jpg/}}
\fi

\begin{document}

\title{Fundamental energy cost of finite-time computing}
\author{Michael Konopik}
 \affiliation{Institute for Theoretical Physics I, University of Stuttgart, D-70550 Stuttgart, Germany}
\author{Till Korten}
 \affiliation{B CUBE - Center for Molecular Bioengineering, Technische Universit\"at Dresden, \tk{D-01307} Dresden, Germany}
\author{Eric Lutz}
 \affiliation{Institute for Theoretical Physics I, University of Stuttgart, D-70550 Stuttgart, Germany}
\author{Heiner Linke}
\affiliation{NanoLund and Solid State Physics, Lund University, S-22100 Lund, Sweden}

\begin{abstract} 
The fundamental energy cost of irreversible computing is given by the Landauer bound of \tk{$kT \ln2$~/bit}. However, this limit is only achievable for infinite-time processes. We here determine the fundamental energy cost of finite-time irreversible computing \er{within  the framework of nonequilibrium thermodynamics}. Comparing the lower bounds of energy required by ideal serial and parallel computers to solve a problem of a given size in a given finite time, we find that the energy cost of a serial computer fundamentally diverges with increasing problem size, whereas that of a parallel computer can stay close to the Landauer limit. We discuss the implications of this result in the context of current technology, and for different degrees of parallelization and amounts of overhead. Our findings provide a physical basis for the design of energy efficient computers.
\end{abstract}

\maketitle

There is wide agreement that Moore's law regarding the exponential growth of the number of components in integrated circuits \cite{Moore:1965} is coming to an end \cite{Theis:2017, Waldrop:2016}. One of the main physical reasons that  prevents further miniaturization  is unavoidable heat generation \cite{Theis:2017, Waldrop:2016}.
A much-improved energy efficiency of computing is therefore a key requirement for any 'More-than-Moore' technology \cite{Arden:2015}.
The fundamental limits to the work cost and the heat dissipation of computing are given by the Landauer bound of $kT\ln 2$ \mk{(in Joule)} per logically irreversible bit operation \cite{Landauer:1961}, where $k$ denotes the Boltzmann constant and $T$ the temperature. The existence of such a lower limit has been established in a number of \er{classical} experiments, using an optical tweezer \cite{ber12}, an electrical circuit \cite{orl12}, a feedback trap \cite{jun14} and nanomagnets \cite{Martini:2016,Hong:2016}, \er{as well as in quantum experiments with a trapped ultracold ion \cite{yan18} and a molecular nanomagnet \cite{gau18}}. However, the Landauer bound is only asymptotically reachable for quasistatic, that is, for infinitely slow processes \cite{lut15}. {In reality, however,} all computing tasks take place in finite time.

We here seek the fundamental minimal energy cost of finite-time computing \er{using the tools of nonequilibrium thermodynamics}. Our aim is to complement discussions of ultimate limits, which, while essential, possess only little practical relevance \cite{Lloyd:2000} or {do not address the fundamental advantages of parallel computing \cite{Meindl:2001}}, and of more applied considerations  \cite{Horvath:2007,Cho:2008}, with only restricted generality.  {Specifically, we} extend the standard Landauer bound by accounting for the nonequilibrium entropy dissipated during the finite-time process close to equilibrium \cite{leb08}. We use the latter to determine the generic thermodynamic limits of parallel and serial finite-time computing \cite{pac07}. \er{Indeed, because of rapid growth in power consumption with increasing processor frequency, \mk{recent} performance gains have been achieved mainly by increased parallelization rather than increased processor frequency \cite{LeSueur2010}. So far, this development is supported by experimental findings \cite{Samani2018} but, to our knowledge, the fundamental limits of the energy consumption of finite-time serial and parallel computation have not been investigated.} We analyze, in particular, how these limits affect the ability of idealized serial and parallel computers to solve a problem of a given size in a given finite time. {A key \mk{insight} is that the energy cost per operation of a parallel computer can be kept close to the Landauer limit even for large problem sizes, whereas that of a serial computer fundamentally diverges. We discuss our results qualitatively for different degrees of parallelization and amounts of overhead operations, and place them quantitatively into the context of existing and emerging technologies.}

\begin{figure}[t]
\includegraphics[width=0.43\textwidth]{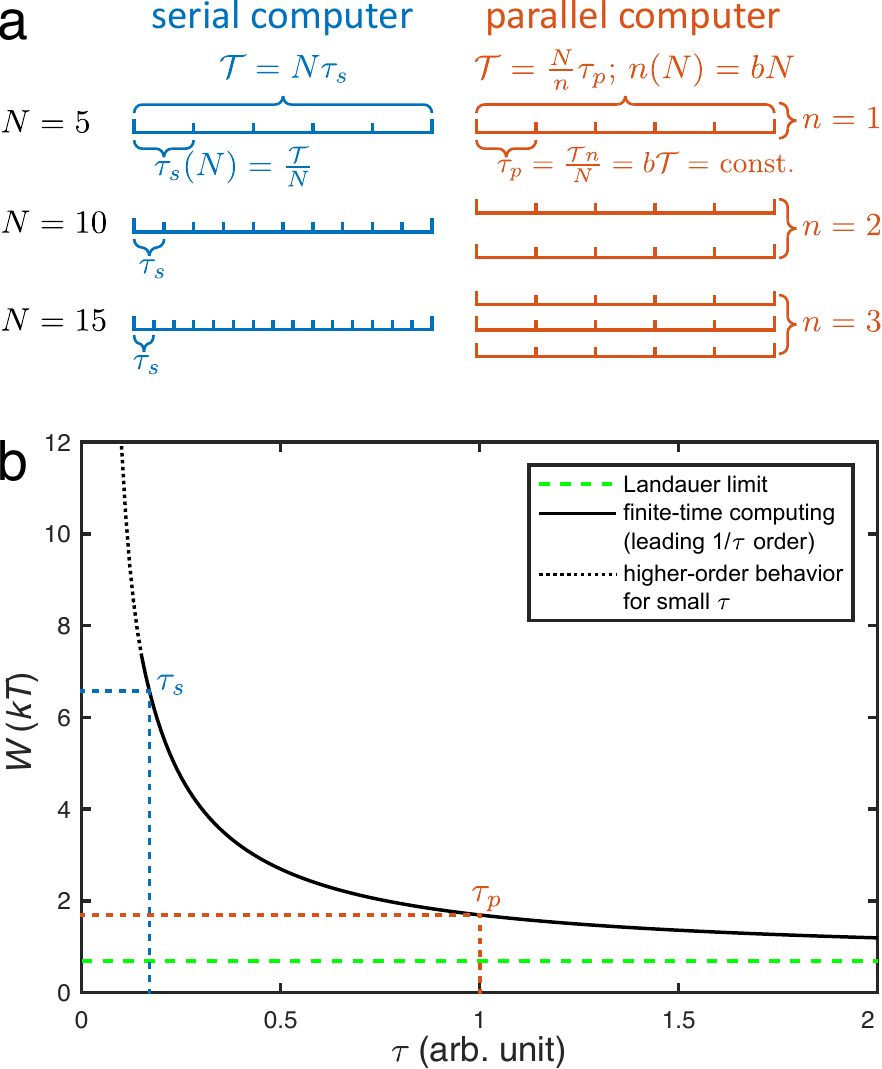}
\caption{{Assumptions for ideal serial and  parallel computers}.  a) Schematic illustration of the three main assumptions: (i) The total time $\mathcal{T}$ available to solve a given problem requiring $N$ computing operations is limited; (ii) an ideal  serial computer (left, blue) reduces the time per operation $\tau_\text{s}$ with increasing problem size $N$; (iii) an ideal  parallel computer (right, red) increases the number of processors $n$ proportional to the size  $N$ in order to keep the time per operation $\tau_\text{p}$ constant. b) Leading $1/\tau$ behavior \tk{(solid line)} of the energy consumption of a single operation of duration $\tau$ near equilibrium, Eq.~(1), and its effects  for serial (blue) and parallel (red) computers. \tk{Higher-order behavior for small $\tau$ (dotted line) is discussed in the Supplemental Material \cite{sm}.}} 
\label{fig_1}
\end{figure}

We base our analysis on the following assumptions (Fig.~\ref{fig_1}): (i) A  computing problem of size $N$ should be solved in {finite time} {$\mathcal{T}$}. In order to stay within this time limit, (ii) an ideal {serial computer dynamically adapts its processing frequency} (time per operation $\tau_\text{s}$; Fig.~\ref{fig_1}a~left), {whereas} (iii) an ideal {parallel computer adapts the number $n$ of processors}{, keeping constant its processing frequency (time per operation $\tau_\text{p}$;  Fig.~\ref{fig_1}a~right)}. We argue that these assumptions are well justified: 
While the available time is not exactly fixed, there is usually a limit on how long calculations can be run. {For example, to be useful, the} weather forecast
for the next day should not run more than a few hours. Moreover, modern processors
implement a number of mechanisms, such as dynamic frequency and voltage scaling \cite{Horvath:2007,Cho:2008}, as well as deactivation of cores using deep-sleep  states \cite{Rotem:2012}, that make them behave in a manner very similar to the assumptions (i)-(iii) made above.

Let us {first} consider a single computation of finite duration $\tau$. Because it occurs in finite time, such {a} nonequilibrium process is necessarily accompanied by the dissipation of an amount of work $W_\text{dis}$  into the environment \cite{leb08}. The energetic cost of a {finite-time}, logically irreversible bit operation may hence be written as a generalized  Landauer bound, 
\begin{equation}
W(\tau) =  k T\ln 2 + W_\text{dis}(\tau),
\label{genLan}
\end{equation}
where $W_\text{dis}/T\geq 0$ is the nonequilibrium entropy produced during the process \cite{leb08}. Close to  equilibrium, the dissipated work is predicted to take the generic form 
$W_\text{dis} = {a}/{\tau}$
(Fig.~\ref{fig_1}b), \er{for both classical \cite{Weidlich:1980,Sekimoto:1997,aur11,pro20}} and quantum \cite{Cavina:2017} dynamics, where $a$ is an \textit{energy efficiency constant} that depends on the system. \er{Interestingly, such a $1/\tau$ behavior \mk{has been observed in experiments}~\cite{ber12,jun14,ma20}}. Equation \eqref{genLan} reduces to the Landauer limit for slow computation processes $\tau \rightarrow \infty$ {and illustrates that in general more work per operation} is required for fast operations and, in turn, more heat is dissipated.

\begin{figure}[t]
\includegraphics[width=0.43\textwidth]{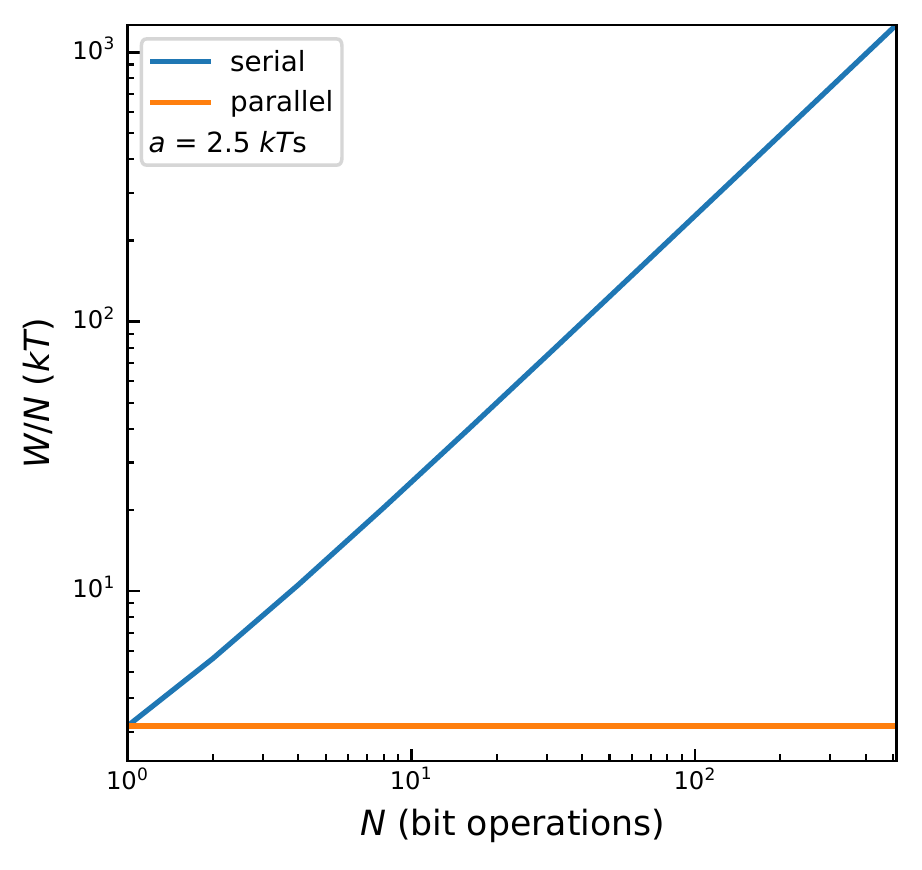}
\caption{Finite-time Landauer bound for ideal  serial and parallel computers.  Energy consumption per operation, $W/N$, for solving a fully parallelizable problem of  size $N$ by an ideal serial, Eq.~(3) (blue), and parallel, Eq.~(4) (orange), computer. The energetic cost diverges with $N$ for an ideal serial computer and remains constant for an ideal parallel computer. Parameters are $\mathcal{T} = 1$~s, $T = 300$~K, $b=1$ and $a = 2.5~kT$s.} 
\label{fig_2}
\end{figure}

In view of Eq.~\eqref{genLan}, the total work cost associated with the solution of  a computing problem that requires $N$  bit operations within the finite time {$\mathcal{T}$} is   given by,
\begin{equation}
W_\text{tot}(N, \tau) = N W(\tau) = N kT\ln 2 + N W_\text{dis}(\tau),
\label{initialW}
\end{equation}
where {$\tau=\tau(\mathcal{T})$} is in general a function of {$\mathcal{T}$}.
 The scaling of the dissipative term with the system size $N$ depends on the type of computing considered. It may be concretely determined for the two idealized computer models introduced above: (i) for an ideal serial computer, the available time per operation decreases with the problem size  as {$\tau_\text{s} = {\mathcal{T}}/{N} =\vcentcolon  1/f_{\text{op}}$} (Fig.~\ref{fig_1}a~left), whereas (ii) for an ideal parallel computer that solves the problem with a number of processors $n(N) = b N$ (with $b\in ]0,1]$; in the following we set $b = 1$) that scales linearly with $N$, the time per operation 
stays constant, {$\tau_\text{p} =  {n \mathcal{T}}/{N} = b\mathcal{T}$} (Fig.~\ref{fig_1}a~right). {The quantity $f_{\text{op}}$ can be interpreted as the operation frequency of the serial processor. }The total energy cost per operation for the serial implementation therefore scales with the system size as,
\begin{equation}
\frac{W_\text{tot}^\text{ser}(N, \mathcal{T})}{N} = kT \ln 2 + \frac{a}{\mathcal{T}}N =kT \ln 2 + a f_{\text{op}}.
\label{serialW}
\end{equation}
The corresponding scaling for the parallel implementation reads,
\begin{equation}
\frac{W_\text{tot}^\text{par}(N, \mathcal{T})}{N} = k T \ln 2 + \frac{a}{b \mathcal{T}}.
\label{parallelW}
\end{equation}
{Equations \eqref{serialW} and \eqref{parallelW} highlight an important, fundamental difference between serial and parallel computing:  whereas the energy cost per operation for a serial computer necessarily increases linearly with $N$, the energy cost per operation for an ideal parallel computer is independent of $N$ (Fig.~\ref{fig_2}); it depends only on the two 
 constants $a$ and $b$ as well as the chosen $\mathcal{T}$. If the chosen computation task permits to choose a large $\mathcal{T}$, then the finite-time energy cost per operation is bounded only by the Landauer limit, even for very large problems $N$.} Equations \eqref{serialW} and \eqref{parallelW} further imply  that
 for a completely parallelizable ideal computer with a  \er{maximum energy consumption} $W_\text{max}$, the maximal problem size $N_\text{max}$ that can be solved within the time limit {$\mathcal{T}$} is proportional to the square root of the  \er{energetic cost}, $\sqrt{W_\text{max}}$, for a serial implementation, whereas it is proportional to the  \er{energetic cost}, $W_\text{max}$,  for a parallel implementation. An ideal parallel computer can therefore, in principle, solve quadratically bigger problems within the same time and  \er{energy} constraints than an ideal serial computer.

{To understand the practical importance of the finite-time energy cost, quantitative values for $a$ are required.  A state-of-the-art general purpose processor that is highly parallel, runs at relatively low clock rate (60 cores \`{a} 4 threads at $f _\text{op} = 1.09$~GHz) and has been thoroughly analyzed for its energy consumption is the Intel Xeon Phi: it consumes $4.5 \cdot 10^{-10} \text{ J/32 bit operation}$ or $a \cdot f_\text{op} = 1.4 \cdot 10^{-11} \text{ J/operation}$ \cite{Shao:2013} (note that this value \mk{accounts} only for computation operations and ignores  more costly transfers to and from memory). This allows us to obtain $a = 1.0 \cdot 10^{-20} \text{ Js} \approx 2.5 
~kT$s \er{(at room temperature; $T=300$ K)} as an estimate for electronic computers. This implies that the finite-time energy cost of an electronic computer exceeds the (quasistatic)  Landauer limit already at a few Hertz of operation frequency. }

\begin{figure}[t]
\vspace{-0.1cm}
\includegraphics[width=0.435\textwidth]{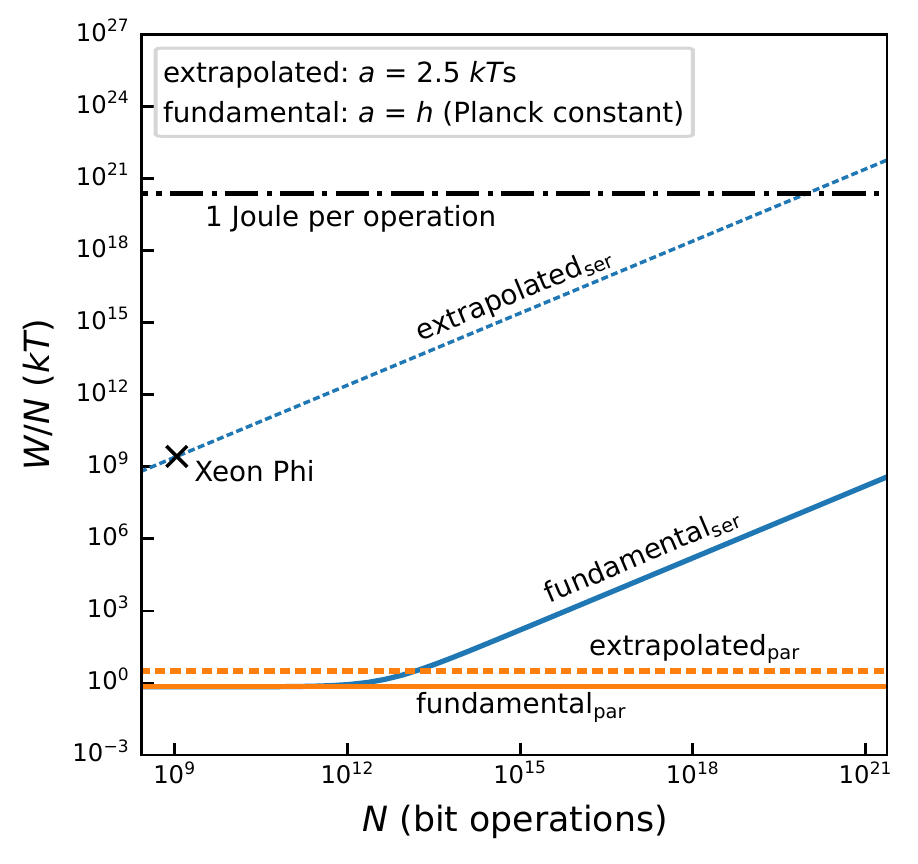}
\caption{Fundamental limit and extrapolated energy cost per operation for ideal serial and parallel computers.  Fundamental limits obtained for $a=h$ (Planck constant) (\tk{solid} lines) and extrapolated energy cost corresponding to $a = 2.5~kT$s (Xeon Phi) (dashed lines) for ideal serial, Eq.~(3) (blue), and parallel, Eq.~(4) (orange), computers.  The measured value \tk{for} a Xeon Phi processor \tk{\cite{Shao:2013}} is represented by a black X. For reference, an energy cost of $1 $ J/operation is shown as a dash-dotted line. Same parameters as in Fig.~\ref{fig_2}.} 
\label{fig_3}
\end{figure}

{Fundamentally, one may argue that the lowest possible value for $a$ is quantum mechanically given by Planck's constant, $h  = 6.6 10^{-34}$ Js $\approx 1.6 \cdot 10^{-13} kT$s \tk{(at room temperature)} \cite{Meindl:2001}}, {13 orders of magnitude lower than the above value for current electronic computers {(Fig.~\ref{fig_3}, solid lines)}.  However, to the best of our knowledge, no physical system has been proposed that would reach such a small value for $a$. In \er{recent} \mk{experimental} studies of the thermodynamics of finite-time operations, much higher values have been found. The lowest \mk{measured} value known to us is $a = 1.1 \cdot 10^{-29}$ Js, reported  for memory operations using molecular nanomagnets \cite{gau18}, corresponding to $a \approx 10^{-6}~kT$s at the operation temperature of $T \approx1$ K. For comparison, from experiments with optical traps  one can estimate $a \approx 2~kT\textrm{s} = 8\cdot  10^{-22}$ Js at \tcb{room temperature} \cite{pro20}.} 

Based on these insights it is illustrative to compare the fundamental energy cost of finite-time computing as a function of problem size $N$ for fully serial and fully parallel computers (Fig.~\ref{fig_3}). For a serial, electronic computer (blue dashed  line) with representative $a = 2.5 ~kT$s, the finite-time energy cost is dominated by the term ${a}N/{\mathcal{T}}$ in Eq.~\eqref{serialW}. A further increase in $N$ (corresponding to an increase in operation frequency $f_{\text{op}}N/ \mathcal{T}$ of a serial computer beyond the currently typical $f_{\text{op}} \approx 1$ GHz) thus leads to a continued increase in energy dissipation per operation. Given that thermal management is already now the limiting factor in processor design, this is not an option unless $a$ can be lowered, for example through transistor and circuit design. If, on the other hand, the quantum mechanical limit of $a \approx h$ were achievable for a serial computer, then the term ${a}N/{\mathcal{T}}$ in Eq.~\eqref{serialW} would become noticeable, compared to the frequency independent  Landauer limit, only once $N$ exceeds $10^{13}$ operations, corresponding to $f_{\text{op}} \approx$  THz (blue line). By contrast, a fully (ideal) parallel computer does not increase its energy cost per operation (orange lines). For $a = 2.5 ~kT$s (Xeon Phi) and $\tau_p = 1$ s, the extrapolated energy cost per operation (orange dashed line) is only about one order of magnitude larger than the fundamental Landauer bound (orange solid line).

\begin{figure}[t]
\includegraphics[width=0.41\textwidth]{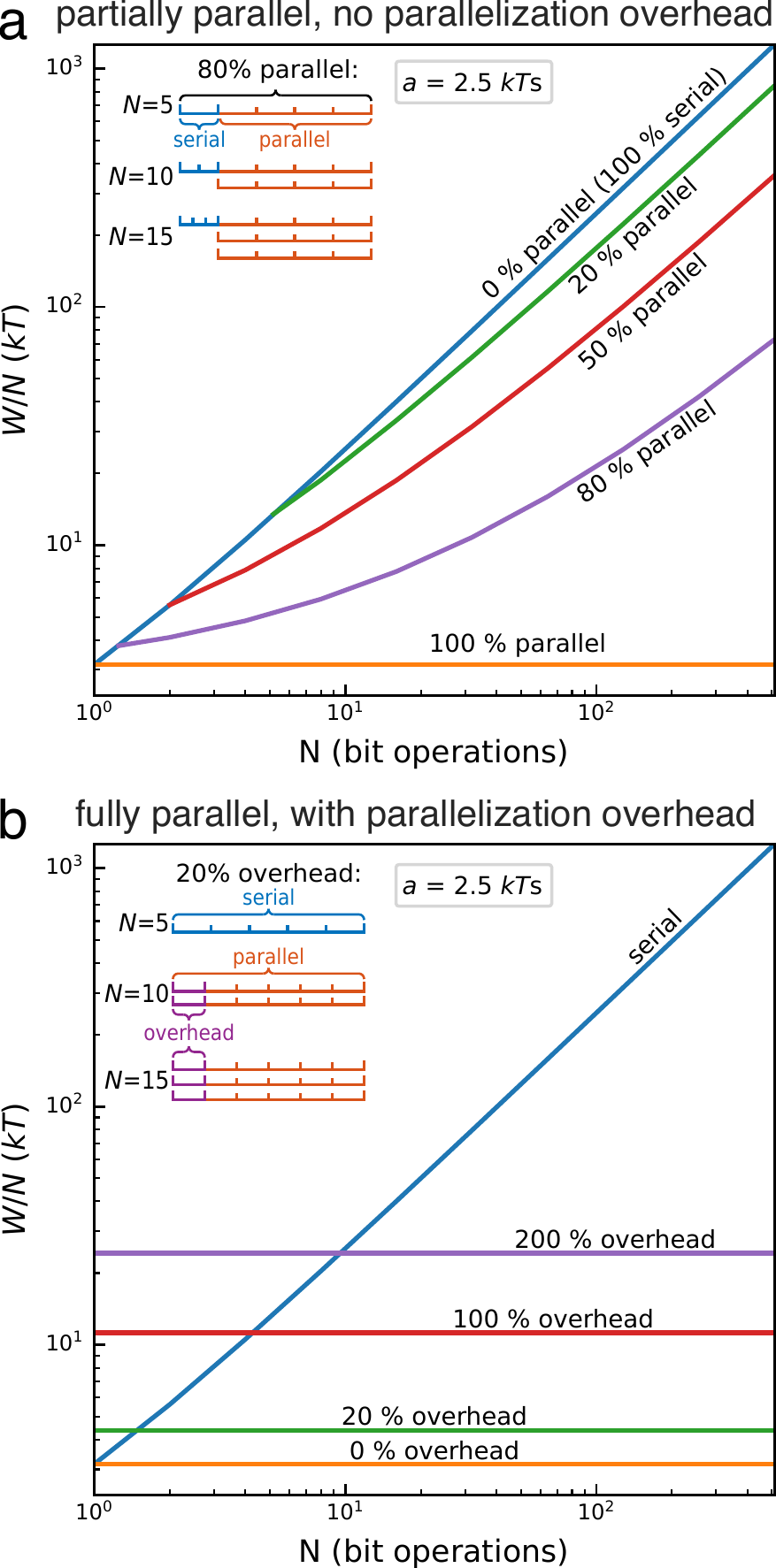}
\caption{Effects of not ideally parallelizable problems.  a) Energy cost per operation for a partially parallelizable algorithm that has no overhead, Eq.~\eqref{amdahlW}. Plots containing a parallel component start at the point where the parallel processor uses one full processing core. b) Energy cost per operation for a fully parallel algorithm with  linear overhead, Eq.~\eqref{oveW}.  Same parameters as in Fig.~\ref{fig_2}. }
\label{fig_4}
\end{figure}

However, real-world algorithms are rarely completely parallelizable\tk{. Therefore,} the \er{ideal} estimates, \er{Eqs.~(3) and (4)}, need to be refined. The impact of non-parallelizable parts of an algorithm on the speedup of parallel computing is commonly described by Amdahl's law \cite{pac07}. According to Amdahl \cite{Amdahl:1967},  the time of the initial serial realization $\mathcal{T}$ can be split into two contributions, a purely serial part $s$, that cannot be done by more than one processor at a time, and a parallel part $p$ that can, ideally, be split equally among all the used $n$ processors (Fig.~\ref{fig_4}a inset).  We evaluate the energetic consequences of such a splitting for our ideal computers \tk{as follows:} We assume that a given problem of size $N$ is comprised of a serial and parallel part, $N=N_\text{s}+N_\text{p}= sN + pN$.  The total computation time is  given by the sum of these two parts,  $\mathcal{T}=\mathcal{T}_\text{s}+\mathcal{T}_\text{p}$, where the serial part $\mathcal{T}_\text{s}$ can be tuned by adjusting the time per operation $\tau_\text{s}$ and the parallel part $\mathcal{T}_\text{p}$ is solely controlled by the number of processors $n$. We then  optimize the combined energy cost function over $\mathcal{T}_\text{p}$ using the fixed total time constraint and obtain   the minimal energy cost for partial parallelization \cite{sm},
\begin{equation}
\frac{W_{\text{tot}}^{\text{com}}}{N} =  kT \ln 2 +  \frac{a }{b  \mathcal{T}}\left(s \sqrt{b N}+\sqrt{p}\right)^2.
\label{amdahlW}
\end{equation}
 It interpolates between the purely serial implementation \eqref{serialW} ($p=0$) and the completely parallelizable processor \eqref{parallelW} ($s=0$). In particular, we observe that the quadratic energetic advantage of the parallel computer is weakened when the degree of parallelization is decreased (Fig.~\ref{fig_4}a).

Another important aspect of real-world algorithms, that ought to be accounted for, is that of parallelization overhead.
Parallelization indeed frequently requires the execution of additional overhead operations $N_\text{ove}$. Usually, this overhead is a function of the number of processors used \cite{pac07}. For concreteness and simplicity, we consider  a linear overhead, $N_\text{ove}(n) = c n$, that corresponds to the case where each processor requires a constant number of overhead operations \er{(different overheads may be  considered)}.
Because of the constant $\mathcal{T}$ assumption, the overhead either means  that each processor needs to work faster in order to compensate  for the overhead (Fig.~\ref{fig_4}b inset), or that one might use a stronger degree of parallelization $1>b' >b$. We  shall assume that   the maximal available parallelization is already used and that  overhead may only be compensated by  adjusting the calculation speed $\tau_\text{p}$.  We then obtain \cite{sm},
\begin{equation}
\tau_\text{p}^\text{ove} = \frac{ \tau_\text{p}}{1+ N_\text{ove}(n)/N}= \frac{b \mathcal{T}}{1+ b c}.
\label{tauPO}
\end{equation} 
Owing to the time dependence of the dissipated work in Eq.~\eqref{genLan}, the energy cost of the parallel execution not only increases with the number of  additional operations $N_\text{ove}$ but also because of the necessary increase in processing speed. As a result, we obtain from Eq.~\eqref{genLan} the total energetic cost in the presence of a linear overhead \cite{sm},
\begin{eqnarray}
\frac{W_\text{tot}^\text{ove}(N)}{N} & =& [N+N_\text{ove}(n)] \frac{W(\tau_\text{p}^\text{ove})}{N} = \left( 1+ \frac{N_\text{ove}(n)}{N}\right) \nonumber \\ 
& &\times  \left[k T \ln 2 + \frac{a}{b \mathcal{T}} \left(1+ \frac{N_\text{ove} (n)}{N}\right)\right].
\label{oveW}
\end{eqnarray}
The  overhead thus causes the parallel computer to be less efficient than the serial computer for small problem sizes. This  is because the Landauer part adds a fixed cost to Eq.~\eqref{oveW}, while the dissipative part will only be dominant  for large $N$. However, the parallel implementation exhibits better scaling and  becomes  more energy efficient for larger problem sizes, even for large overhead (Fig.~\ref{fig_4}b). This result holds true as long as $N_\text{ove}(n)$ scales better than $n^{3/2}$ (or, equivalently, $N^{3/2}$, since \tcb{here}  $n\propto N$). 

In conclusion, we have used insights from nonequilibrium thermodynamics to develop a general formalism to evaluate  the fundamental energetic cost of finite-time computing. Our main result is that the finite-time energy cost per operation of a fully parallel computer is independent of problem size and can realistically operate close to the Landauer limit. This is in contrast to serial computers for which the finite-time energy cost per operation necessarily increases with problem size.  We also provide a framework for including partial parallelization and the associated energy overhead into the analysis. For serial computers, the key limiting factor is the finite-time constant $a$. To enable a drastic increase in operation frequency without fundamentally necessary, prohibitive energy consumption,  $a$ needs to be strongly reduced  below its current value of  $a \approx kT$s in electronic computers. Whether physical systems are available that allow $a$ smaller than the current record of $a\approx  10^{-6}~kT$s \cite{gau18}, possibly down to a quantum limit of $a\approx  h$, is an open question. At the same time, massively parallel computers can be realized in biological computing, such as DNA computing \cite{Adleman:1994,Braich:2002,Erlich:2017} or network-based biocomputing (NBC) \cite{Nicolau:2006,Nicolau:2016}, which use small DNA molecules or cytoskeletal filaments, respectively, as computing cores and memory. These are cheap to mass-produce and can be added to the computation in amounts matching the problem size. Both DNA computing and NBC have been estimated to be able to work very close to the Landauer limit per operation \cite{Nicolau:2016}.  As shown in our study, this finite-time energy cost is independent of problem size (Fig.~3). From the perspective of finite-time energy cost, biological computers thus  offer a potentially large, fundamental advantage over electronic computers.

We acknowledge   financial support from the German Science Foundation (DFG) (Contract No FOR 2724) \tk{and from the European Union's Horizon 2020 research and innovation programme under grant agreement No 732482 (Bio4Comp)}, and from the Knut and Alice Wallenberg Foundation (project 2016.0089).


\bibliographystyle{naturemag}

\bibliography{references}


\section{Supplemental Material}
\setcounter{figure}{0}
\renewcommand{\thefigure}{S\arabic{figure}}
\setcounter{table}{0}
\renewcommand{\thetable}{S\arabic{table}}
\setcounter{equation}{0}
\renewcommand{\theequation}{S\arabic{equation}}

\textit{Partially parallelizable algorithms.} The energy cost of partially parallelizable problems \eqref{amdahlW} is derived from three observations: (i) the problem can be split into a serial and a parallel part, $N= N_\text{s} + N_\text{p}= s N + p N$, (ii) corresponding to a total duration   $\mathcal{T}=\mathcal{T}_\text{s} + \mathcal{T}_\text{p}$, where the respective serial and parallel time allocations, $\mathcal{T}_\text{s}$ and $\mathcal{T}_\text{p}$,  can be freely chosen, except for the fixed total time constraint. Since we are interested in the optimal energy cost, it is important to further (iii) optimize the energy cost function over $\mathcal{T}_\text{p}$, respecting the time constraint. The computation times for a single operation, $\tau_\text{s}$ and $\tau_\text{p}$ (for notational simplicity,  $\tau_\text{s}$ denotes the time for the serial part, not the serial computation) are thus,
\begin{equation}
\begin{aligned}
\mathcal{T}_\text{s} &= \tau_\text{s} N_\text{s} \quad \text{ and } \quad \mathcal{T}_\text{p} = \tau_\text{p} N_\text{p}/n(N_\text{p}).
\end{aligned}
\end{equation}
The total energy cost follows from Eq.~\eqref{genLan} as,
\begin{equation}
\begin{aligned}
W^{\text{com}}_\text{tot} &= N kT \ln 2 + \frac{a}{\tau_\text{s}} N_\text{s} + \frac{a}{\tau_\text{p}} N_\text{p}\\
&= N\left[kT \ln 2 + \frac{s^2 N a}{\mathcal{T}_\text{s}} + \frac{p^2 a N}{\mathcal{T}_\text{p} n(N_\text{p})}\right].
\label{partparrW}
\end{aligned}
\end{equation}
Minimizing  with respect to $\mathcal{T}_\text{p}$ using (iii), we obtain, 
\begin{equation}
\mathcal{T}_\text{p}=\frac{a N [p + \sqrt{n(N_\text{p})} s]^2}{n(N_\text{p}) \mathcal{T}}.
\end{equation}
Inserting this result into \eqref{partparrW}, we then find,
\begin{equation}
W^{\text{com}}_\text{tot}= N\left(kT \ln2 +\frac{a  \left[p + \sqrt{n(N_p)} s\right]^2}{n(N_p) \mathcal{T}}\right).
\end{equation}
Equation \eqref{amdahlW} eventually follows with  $n(N_p)= b N p$.\\

\textit{Parallelization overhead.} There are many possible kinds of overhead. We  consider the  simple linear form, which is linear in the number of processors and thus linear in the problem size \tk{for an ideal parallel computer}. For simplicity, we assume that all available processors are used and therefore $n(N)$ is fixed. The overhead is accordingly  taken into account by modulating the computation speed. With  $N_\text{g} = N + N_\text{ove}(n)$ and  $\tau_\text{p} = N/n(N)$, the general time needed to solve the problem including the linear overhead is,
\begin{equation}
\mathcal{T}'=\tau_\text{p}' \frac{N_\text{g}}{n(N)}= \tau_\text{p}' \left[ \frac{N}{n(N)}+\frac{N_\text{ove}(n)}{n(N)}\right].
\end{equation}
The given time constraint is $\mathcal{T}'=\mathcal{T}= N \tau_\text{p}/n(N)$ for the overhead-free case. We thus have,
\begin{equation}
\tau_\text{p}'= \tau_\text{p} \frac{1}{1+ {N_\text{ove}(n)}/{N}}.
\end{equation}
The total energy cost is finally,
\begin{equation}
W_\text{tot}^\text{ove}= N \left( 1+ \frac{N_\text{ove}(n)}{N}\right) \left[  k T \ln 2 + \frac{a}{b \mathcal{T}} \left(1+ \frac{N_\text{ove}(n)}{N}\right)\right].
\label{s7}
\end{equation}
Equation \eqref{s7} can in principle be extended for any kind of overhead.

\er{\textit{Corrections to the $1/\tau$ behavior.} Equations \eqref{serialW} and \eqref{parallelW} are derived using the leading-order finite-time correction to the quasistatic driving. In general applications, this limit may not be reached and it is thus of interest to consider the influence of  higher orders $W_{\text{irr}}\propto 1/\tau^m$, $m>1$, on the energetic bounds for  serial and parallel computation. Higher orders  lead to serial costs of the form $W_{\text{ser}}/N\propto 1/\tau_s^m=  N^m/\mathcal{T}^m$. As a result, the larger the value of $m$, the worse the energy scaling becomes. On the other hand, in the parallel case we have $W_\text{par}/N\propto1/\tau_p^m = 1/(b\mathcal{T})^m$, with $b\leq1$. Remarkably, the scaling is still independent of the number of operations $N$. Hence, while there is no $N$ dependence in the parallel case, the constant can be quite large compared to the Landauer bound, if $b^m$ is very small. 
All in all, the leading-order comparison of  ideal serial and parallel computation is, in a sense, a favorable comparison for the serial case, as higher order make the advantage of the (ideal) parallel computation more distinct.}

\end{document}